\documentclass[prl,twocolumn,superscriptaddress,showpacs]{revtex4}

\usepackage{graphicx}
\usepackage{verbatim}
\usepackage{color}
\usepackage{amsmath}
\usepackage{microtype}

\begin{document}

\newcommand{\bra}[1]{\left\langle #1\right|}
\newcommand{\ket}[1]{\left|#1\right\rangle}
\newcommand{\braket}[2]{\left\langle #1|#2\right\rangle}
\newcommand{\com}[2]{\left[#1,#2\right]}
\newcommand{\braketop}[3]{\left\langle #1\left|#2\right|#3\right\rangle}
\newcommand{\mean}[1]{\left\langle #1 \right\rangle}
\newcommand{\trace}[2][]{{\rm Tr_{#1}}\left(#2\right)}
\newcommand{\ImaginaryPart}{{\rm Im}}
\newcommand{\RealPart}{{\rm Re}}
\newcommand{\leftexp}[2]{{\vphantom{#2}}^{#1}{#2}}
\newcommand{\leftind}[2]{{\vphantom{#2}}_{#1}{#2}}
\newcommand{\elem}{\in}
\newcommand{\rp}{\right)}
\newcommand{\lp}{\left(}
\newcommand{\lcb}{\left\{}
\newcommand{\rcb}{\right\}}
\newcommand{\rsb}{\right]}
\newcommand{\lsb}{\left[}
\newcommand{\lbv}{\left|}
\newcommand{\rbv}{\right|}
\newcommand{\lvb}{\lbv}
\newcommand{\rvb}{\rbv}
\newcommand{\bs}{\boldsymbol}
\renewcommand{\inf}{\infty}
\newcommand{\myfrac}[2]{^{#1\negthickspace\negthickspace}/_{\negthinspace#2}}
\newcommand{\mycaption}[2]{\caption[#1]{\small #1 #2}}
\newcommand{\order}[1]{{{\mathcal O}\lp#1\rp}}
\newcommand{\iohbar}{\frac{-i}{\hbar}}
\newcommand{\melem}[1]{_{#1}}

\newcommand{\qed}{EDQ}
\newcommand{\laqed}{l'\qed}
\newcommand{\Laqed}{L'\qed}
\newcommand{\etal}{{\it et al.}}
\newcommand{\go}{«~}
\newcommand{\gf}{~»}

\newcommand{\Nphot}{\aleph}
\newcommand{\wa}{{\omega_a}}
\renewcommand{\wr}{{\omega_r}}
\newcommand{\warot}{{\omega_{a{\rm rot}}}}
\newcommand{\wrrot}{{\omega_{r{\rm rot}}}}
\newcommand{\epm}{{\epsilon_m}}
\newcommand{\epc}{{\epsilon_c}}
\newcommand{\wm}{{\omega_m}}
\newcommand{\wc}{{\omega_c}}
\newcommand{\gd}{{\gamma_\downarrow}}
\newcommand{\gu}{{\gamma_\uparrow}}
\newcommand{\gphi}{{\gamma_\varphi}}
\newcommand{\gam}{{\gamma_1}}
\newcommand{\UnitaryMatrix}{{\mathbf 1}}

\newcommand{\e}{e}
\newcommand{\g}{g}
\newcommand{\eg}{\myfrac{\e}{\g}}
\renewcommand{\ge}{\myfrac{\g}{\e}}
\renewcommand{\ae}{\alpha_\e}
\newcommand{\ag}{\alpha_\g}
\newcommand{\aeg}{\alpha_{\e,\g}}
\newcommand{\apm}{\genfrac{}{}{0pt}{1}{\mu}{\beta}}
\newcommand{\am}{\beta}
\renewcommand{\ap}{\mu}
\renewcommand{\ne}{n_\e}
\renewcommand{\ng}{n_\g}
\renewcommand{\neg}{n_{\e,\g}}
\newcommand{\qrho}{{\rho}}
\newcommand{\crho}{{\varrho}}

\newcommand{\ad}{{a^\dag}}
\renewcommand{\sp}{\sigma_+}
\newcommand{\sm}{\sigma_-}
\newcommand{\spm}{\sigma_\pm}
\newcommand{\sx}{\sigma_x}
\newcommand{\sy}{\sigma_y}
\newcommand{\sz}{\sigma_z}
\newcommand{\si}{\sigma_i}
\newcommand{\ip}{{I_+}}
\newcommand{\im}{{I_-}}
\newcommand{\ipm}{{I_\pm}}
\newcommand{\imp}{{I_\mp}}
\newcommand{\ada}{{\ad a}}
\newcommand{\Pe}{\Pi_\e}
\newcommand{\Pg}{\Pi_\g}
\newcommand{\Peg}{\Pi_{\e,\g}}
\newcommand{\Pa}{\Pi_\alpha}
\newcommand{\Dop}{D}

\newcommand{\superop}[1]{{\mathcal #1}}
\newcommand{\sD}{{\superop{D}}}
\newcommand{\sL}{{\superop{L}}}
\newcommand{\sC}{{\superop{C}}}
\newcommand{\sT}{{\superop{T}}}
\newcommand{\sM}{{\superop{M}}}

\newcommand{\tr}[1]{\mathbf{#1}}
\newcommand{\tU}{{\tr{U}}}
\newcommand{\tS}{{\tr{S}}}
\newcommand{\tD}{{\tr{D}}}
\newcommand{\tR}{{\tr{R}}}
\newcommand{\tP}{{\tr{P}}}
\newcommand{\tT}{{\tr{T}}}
\newcommand{\trans}[1]{^{#1}}

\newcommand{\atr}[1]{{\mathbf #1}}
\newcommand{\atU}{{\atr{U}}}
\newcommand{\atS}{{\atr{S}}}
\newcommand{\atD}{{\atr{D}}}
\newcommand{\atR}{{\atr{R}}}
\newcommand{\atP}{{\atr{P}}}
\newcommand{\atT}{{\atr{T}}}

\newcommand{\aop}{a}
\newcommand{\eq}[1]{Eq.~(\ref{#1})}

\newcommand{\nn}{\nonumber}
\newcommand{\nl}{\nn \\ &&}
\newcommand{\dg}{^\dagger}
\newcommand{\rt}[1]{\sqrt{#1}}
\renewcommand{\vec}[1]{\underset{\widetilde{}}{#1}}
\newcommand{\smallfrac}[2]{\mbox{$\frac{#1}{#2}$}}
\newcommand{\ito}{It\^o~}
\newcommand{\str}{Stratonovich~}
\newcommand{\sch}{Schr\"odinger~}
\newcommand{\schs}{Schr\"odinger's~}
\newcommand{\erf}[1]{Eq.~(\ref{#1})}
\newcommand{\erfs}[2]{Eqs.~(\ref{#1}) and (\ref{#2})}
\newcommand{\erft}[2]{Eqs.~(\ref{#1}) -- (\ref{#2})}
\newcommand{\szo}{\hat{\sigma}^z}
\newcommand{\sxo}{\hat{\sigma}^x}
\newcommand{\syo}{\hat{\sigma}^y}
\newcommand{\smo}{\hat{\sigma}^-}
\newcommand{\spo}{\hat{\sigma}^+}
\newcommand{\ano}{\hat{a}}
\newcommand{\cro}{\hat{a}\dg}
\newcommand{\Ho}{\hat{H}}
\newcommand{\Mso}[2]{{\cal M}_{#2}[{#1}]}

\newcommand{\red}{\color[rgb]{0.8,0,0}}
\newcommand{\green}{\color[rgb]{0.0,0.6,0.0}}
\newcommand{\dkgrn}{\color[rgb]{0.0,0.4,0.0}} 
\newcommand{\blu}{\color[rgb]{0,0,0.6}}
\newcommand{\blue}{\color[rgb]{0,0,0.6}}
\newcommand{\pur}{\color[rgb]{0.8,0,0.8}}
\newcommand{\blk}{\color{black}}

\newcommand{\dage}{\dot{\alpha}_{\g,\e}}
\newcommand{\age}{\alpha_{\g,\e}}
\newcommand{\dge}{\delta_{\g,\e}}
\newcommand{\ppm}{\stackrel{\raisebox{-1pt}[0pt][0pt]{$\scriptstyle(+)$}}{\raisebox{-4pt}[0pt][0pt]{$-$}}}

\title{Non-linear dispersive regime of cavity QED: The dressed dephasing model}
\date{\today}

\author{Maxime Boissonneault}
\affiliation{D\'epartement de Physique et Regroupement Qu\'eb\'ecois sur les Mat\'eriaux de Pointe, Universit\'e
de Sherbrooke, Sherbrooke, Qu\'ebec, Canada, J1K 2R1}
\author{Jay Gambetta}
\affiliation{Institute for Quantum Computing and Department of Physics and Astronomy, University of Waterloo, Waterloo, Ontario N2L 3G1, Canada}
\affiliation{Centre for Quantum Computer Technology, Centre for Quantum Dynamics, Griffith University, Brisbane, Queensland 4111, Australia}
\author{Alexandre Blais}
\affiliation{D\'epartement de Physique et Regroupement Qu\'eb\'ecois sur les Mat\'eriaux de Pointe, Universit\'e
de Sherbrooke, Sherbrooke, Qu\'ebec, Canada, J1K 2R1}

\begin{abstract}
	Systems in the dispersive regime of cavity quantum electrodynamics (QED) are approaching the limits of validity of the dispersive approximation.  We present a model which takes into account non-linear corrections to dressing of the atom by the field. We find that in presence of pure dephasing, photons populating the cavity act as a heat bath on the atom, inducing incoherent relaxation and excitation. These effects are shown to reduce the achievable signal-to-noise ratio in cavity QED realizations where the atom is measure indirectly through cavity transmission.	
\end{abstract}


\maketitle


Cavity quantum electrodynamics (QED) studies the coupling of a two-level system (TLS) to one mode of a quantized light field.  This is traditionally realized in systems where the TLSs are Rydberg~\cite{haroche:2006a,gleyzes:2007a,guerlin:2007a} or alkali~\cite{boca:2004a,maunz:2005a,brennecke:2007a} atoms and the light field enclosed in a high-finesse cavity.  Recently, this active field of research has attracted even more attention due to the realization of cavity QED in solid-state systems.  Examples are semiconducting~\cite{hennessy:2007a,englund:2007a} and superconducting systems~\cite{wallraff:2004a,chiorescu:2004a,johansson:2006,sillanpaa:2007a} where the strong coupling regime was achieved.  Mechanical oscillators could also be used to reach this regime~\cite{armour:2002a,irish:2003a}.

In this letter, we study the dispersive regime of cavity QED, where the TLS-cavity detuning is larger than the coupling strength.  With the recent realization of the novel {\em strong} dispersive limit in atomic~\cite{gleyzes:2007a,guerlin:2007a}, superconducting~\cite{schuster:2007a} and micromechanical~\cite{thompson:2008a} systems, this regime offers new possibilities to study light-matter interaction at its most fundamental level.  We study the dispersive limit by going beyond the usual linear approximation~\cite{haroche:2006a,blais:2004a} and take into account important non-linear contributions.  We obtain a reduced master equation (ME) and quantum trajectory equation describing the TLS.  We show that, due to dressing of the TLS by the field, pure TLS dephasing causes photons in the cavity to act as a heat bath on the TLS.  Using these results, we obtain an expression for the signal-to-noise ratio (SNR) in a homodyne measurement of the TLS state.  Contrary to initial expectations~\cite{blais:2004a,gambetta:2008a}, this SNR saturates with increased measurement power.  This result is consistent with experimental results obtained in a cavity QED realization based on superconducting circuits (circuit QED)~\cite{houck_private}.  Based on this observation, we suggest a path to optimize the SNR of dispersive measurements.  Our work can be applied to all cavity QED realizations reaching the strong coupling regime and will help in increasing the measurement fidelity~\cite{gambetta:2007a} of any TLS dispersively coupled to a harmonic oscillator.  These results could, for example, be applied to superconducting qubit architectures which are dispersively measured~\cite{boulant:2007a,lupascu:2007a}.


Cavity QED is described by the Jaynes-Cummings Hamiltonian~\cite{haroche:2006a}
\begin{equation}\label{eq:HJC}
H_s =  \hbar\wr \ad\aop + \hbar\wa \frac{\sz}{2} + \hbar g\left( \ad \sm + a\sp \right),
\end{equation}
where $a^{(\dag)}$ and $\spm$ are ladder operators for the photon field and the TLS respectively, $\wr$ the cavity frequency, $\wa$ the TLS transition frequency and $g$ their coupling strength.  We are interested in the dynamics of the TLS in the presence of photon population of the resonator.   For this purpose, a coherent drive on the cavity is modeled by the Hamiltonian
	$H_d  = \hbar \epsilon_m(t) \left(  a^\dag e^{-i\omega_m t} + a e^{i\omega_m t} \right)$,
where $\epsilon_m(t)$ and $\omega_m$ denote measurement amplitude and frequency.


Energy relaxation results from the coupling of the cavity and TLS to independent baths of harmonic oscillators~\cite{carmichael:1993a},
\begin{equation}
	\label{eq:bathcoupling}
		H_{j} = i\hbar\int_{0}^\inf \sqrt{g_j(\omega)} \left[ b_j^\dag(\omega) - b_j(\omega)\right] \left[ c_j^\dag+c_j\right] d\omega,
\end{equation}
where $j=\kappa$ or $\gamma$ represent either the cavity or the TLS bath with $c_\kappa = a$ and $c_\gamma = \sigma_-$. Here, $g_j(\omega)$ is the coupling strength to the bath mode of frequency $\omega$.  In the Born-Markov approximation, integrating out the baths leads to a TLS-cavity ME of Lindblad form with cavity decay rate $\kappa = 2\pi |g_\kappa(\omega_r)|^2$ and TLS relaxation rate $\gamma_1 = T_1^{-1}= 2\pi |g_\gamma(\omega_a)|^2$.  TLS dephasing can be modeled by coupling to a longitudinal classical fluctuating parameter $f(t)$ using the Hamiltonian
	$H_\varphi = \hbar \upsilon f(t)\sz,$
where $\upsilon$ characterizes the coupling strength of the TLS to the fluctuations.  This leads to pure dephasing of the TLS at a rate $\gamma_\varphi = 2 \upsilon^2S_f(\omega\rightarrow0)$, with $S_f(\omega)$ the noise spectrum of $f(t)$.


The dispersive regime of cavity QED is realized when  $|\Delta|\equiv |\wa - \wr| \gg g$. In this situation, it is useful to move to the dispersive basis by using a unitary transformation to diagonalize \eq{eq:HJC}.  This is done using 
	$\tD = \exp\left[{\Lambda(N_q) (a^\dag\sigma_--a\sigma_+)}\right]$,
where $\Lambda(N_q) = \arctan\lp2\lambda\sqrt{N_q}\rp/2\sqrt{N_q}$ and $N_q \equiv \ad\aop + \ket{e}\bra{e}$ is an operator representing the total number of excitations. While this can be done exactly, here we only present the result to third order in the small parameter $\lambda=g/\Delta$
\begin{equation}
	\label{eqn:DispersiveHamiltonianCompact}
	\begin{split}
		\raisetag{36pt}
		H_s\trans{\atD} &= \tD^\dag H_s \tD\\ 
		& \approx \hbar\lsb\wr+\zeta+(\chi+\zeta\ada)\sz\rsb\ada + \hbar(\wa+\chi)\frac{\sz}{2},
	\end{split}
\end{equation}
where $2\chi = 2g^2(1-\lambda^2)/\Delta$ is the Stark shift per photon.  This Hamiltonian is similar to the usual dispersive Hamiltonian~\cite{haroche:2006a,blais:2004a}, but the third-order expansion leads to corrections to $\chi$, $\omega_r$, and also yields a squeezing term $(\ada)^2$ of amplitude $\zeta = -g^4/\Delta^3$.

\begin{figure}
	\centering
	\includegraphics[width=0.750\hsize]{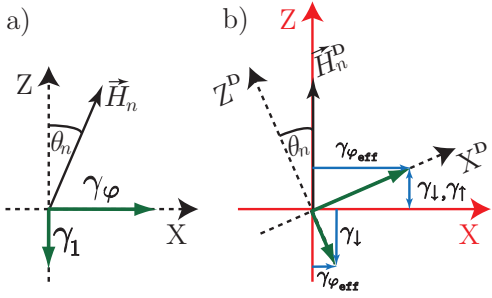}
	\caption{(Color online) Geometrical representation of $H_s$ in the subspace $\mathcal E_n = \{\ket{n,\g},\ket{n-1,\e}\}$  with decay and dephasing rates for a) the bare and b) dispersive basis.  The axes $X\trans{\atD}$ and $Z\trans{\atD}$ are the original axes which are rotated by an angle $\theta_n = \arctan(2\lambda\sqrt{n})$ around the $Y$ axis such that $H_n\trans{\atD}$ is along the $Z$ direction.}
	\label{fig:model_geo}
\end{figure}

The transformation $\tD$ to the dispersive basis can be interpreted as a rotation in the TLS-cavity space.  To understand how this transformation will affect dissipation, it is useful to consider $\mathcal E_n = \{\ket{n,\g},\ket{n-1,\e}\}$, the $n$-excitation subspace.  In $\mathcal E_n$, $H_s$ takes the form	$H_n = \hbar\Delta \bar\sigma_z/2 + \hbar g\sqrt{n} \bar\sigma_x$, where $\bar\sigma_{x,z}$ act as Pauli operators on $\mathcal E_n$. As illustrated in Fig.~\ref{fig:model_geo}, this Hamiltonian is diagonalized by a rotation about the $Y$ axis of angle $\theta_n = \arctan(2\lambda\sqrt{n})$.  The first order dispersive approximation simply corresponds to taking $\theta_n \approx 2\lambda\sqrt{n}$, while the exact transformation $\tD$ corresponds to performing the above rotation in all subspaces $\mathcal E_{n>0}$.  After this transformation, the new eigenstates are entangled superpositions of TLS and cavity states.  Hence, in the dispersive basis, the TLS is dressed by the field and acquires a photon part.

As illustrated in Fig.~\ref{fig:model_geo}, it is also important to rotate the system-bath Hamiltonians.  We represent the TLS relaxation rate $\gamma_1$ by an arrow pointing along the original axis $Z^D$. After rotation, it acquires a component along the $X$ axis, corresponding to dephasing, and the magnitude of its $Z$ component is reduced.  In the same way, we can represent dephasing as an arrow pointing along $X^D$.  After rotation, it acquires a component along the $Z$ axis, corresponding to upward and downward transitions of the TLS, and the magnitude of its $X$ component is reduced.  Since the rotation angle $\theta_n$ varies with $n$,  from this simple picture we expect the effective TLS decay and dressed dephasing rates to depend on the subspace $\mathcal E_n$ and thus on the number of photons in the cavity.

To obtain an effective ME for the TLS in the third-order dispersive approximation, we apply the transformation $\tD$ to $H_s$, $H_\kappa$, $H_\gamma$, $H_\varphi$ and $H_d$. We then integrate out the baths to obtain  Lindblad form dissipators for the TLS and cavity operators~\cite{carmichael:1993a}.  Starting from this transformed ME, our goal is to obtain an effective ME for the TLS only.  To arrive at this result, we follow Ref.~\cite{gambetta:2008a} in tracing out the cavity degrees of freedom by first applying a polaron-type transformation to the TLS-cavity ME.  While in the linear dispersive approximation this results in an exact TLS ME~\cite{gambetta:2008a}, in the non-linear case we approximate the photon number operator $a^\dag a$ by linearizing quantum fluctuations around the classical value.  In this way, we obtain the following ME for the TLS density matrix $\qrho\trans{\atD}$ in the dispersive basis
\begin{equation}
	\label{eqn:ReducedMasterEquation}
	\begin{split}
		\dot\qrho\trans{\atD} 
		 &= -i\frac{\tilde\omega_{a}\trans{\atD}}{2}\com{\sz}{\qrho\trans{\atD}} + \frac{\gphi_{\rm eff}}{2}\sD[\sz]\qrho\trans{\atD} \\
		&\quad  + \gd\sD[\sm]\qrho\trans{\atD} + \gu\sD[\sp]\qrho\trans{\atD}
		\equiv \sL\trans{\atD} \qrho \trans{\atD}.
	\end{split}
\end{equation}
In this expression, $\tilde\omega_{a}\trans{\atD}$ is the Lamb and Stark shifted TLS transition frequency, including contributions both linear~\cite{gambetta:2008a} and quadratic in photon number (i.e.~leading to squeezing).  The effective dephasing rate $\gphi_{\rm eff}$ takes into account measurement-induced dephasing~\cite{gambetta:2006a}, with corrections due to the non-linear terms.  Finally, the effective decay and excitation rates are given by
\begin{equation}
	\label{eqn:Gamma_Up_Down}
	\gd = \gam\lsb1-\frac{2\bar n+1}{4n_\mathrm{crit}}\rsb + \gamma_\kappa + \gamma_{\Delta}\bar n, \quad \gu = \gamma_{-\Delta} \bar n, 	
\end{equation}
where $\gamma_\kappa=2\pi|g_\kappa(\omega_a)|^2/4n_\mathrm{crit}$ is the Purcell decay rate and $\gamma_{\pm\Delta}  = \upsilon^2S_f(\pm\Delta)/n_\mathrm{crit}$ are measurement and dephasing induced excitation and relaxation rates. In these expressions, $n_{\rm crit} = \Delta^2/4g^2$ is the critical photon number~\cite{blais:2004a} and $\bar n = P_\g n_\g + P_\e n_\e$ the average number of photons in the cavity, with $P_{g,e}$ the ground and excited state population and $n_{\g,\e} = |\age|^2$ the photon population corresponding to the classical fields $\age$ associated to the TLS ground and excited states. These satisfy
\begin{equation}
	\label{eqn:Condition_alphas}
	\begin{split}
	\dage &= - i\lsb\Delta_{rm}' \mp \dge\rsb\age -i\epm\left[1\mp \frac{1}{8n_{\rm crit}}\right] \\
		  & \quad - \frac{\kappa}{2}\left[1\mp \frac{1}{4n_{\rm crit}}\right]\age,
	\end{split}
\end{equation}
where $\Delta_{rm}' = \wr-\wm+\zeta$ is the rotating-frame cavity frequency and $\dge = \chi+2\zeta n_{\g,\e}$ the cavity pull.

\begin{figure}[tp]
	\centering
	\includegraphics[width=0.9\hsize]{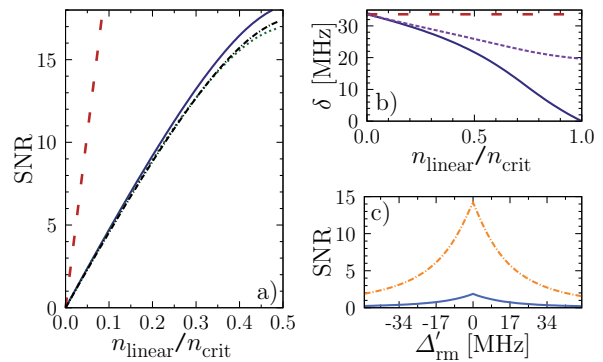}
	\caption{(Color online) a) SNR vs $n_\mathrm{linear}/n_\mathrm{crit}$ for $\Delta_{rm}'=0$ in the linear (dashed red line) and second order (full blue line) models.  Dotted green line is Eq.~(\ref{eq:SNRapprox}). Dashed-dotted black line includes transients.  b) Cavity pull $\delta$ vs $n_\mathrm{linear}/n_\mathrm{crit}$ in the linear model (dashed red line), second order approximation (full blue line) and exact (dotted purple line).  c) SNR vs measurement detuning $\Delta_{rm}'$ at fixed photon population (dashed-dotted orange line $n=10$, full blue line $n=1$).  
	The parameters are $(\Delta,g,\kappa,\gam,\gphi)/2\pi = (1700, 170, 34, 0.1, 0.1)$ MHz, $\gamma_{\pm\Delta}=2\lambda^2\gphi$ and $\eta=1/80$ consistant with experimental observations~\cite{houck_private}. These parameters correspond to $n_\mathrm{crit}=25$. 
	}
	\label{fig:snr}
\end{figure}

As discussed in relation to the geometrical picture of Fig.~\ref{fig:model_geo}, in the dispersive picture, the TLS relaxation rate can be reduced by photon population [negative sign in LHS of \eq{eqn:Gamma_Up_Down}]. Dressing of the TLS with increasing $\bar n$ renders it less sensitive to noise at $\wa$. However, the relaxation rate is now enhanced by noise at the frequency $\Delta$.  In the same way, the TLS acquires a finite excitation rate $\gu$.  Dressing of dephasing thus leads to the measurement photon acting as a heat bath for the TLS.  We have carried out extensive numerical calculations which have shown this effective ME to accurately capture the dynamics of the TLS up to a photon population of the order of $n_{\rm crit}$ for the parameters used in Fig.~\ref{fig:snr}.

The variation of these modified rates with photon population can easily be probed in cavity QED implementations in which  the TLS is measured directly~\cite{gleyzes:2007a,guerlin:2007a,sillanpaa:2007a}.  In many realizations of cavity QED however, measurement of the TLS is realized indirectly by probing the signal transmitted through the cavity in a homodyne measurement~\cite{boca:2004a,maunz:2005a,wallraff:2004a}.   Since the ME description does not take into account the result of the observation, we use quantum trajectory theory~\cite{carmichael:1993a,wiseman:1993a} to include this information and derive the evolution equation for the conditional state.  This was done in Ref.~\cite{gambetta:2008a} for the linear dispersive model and is extended here to incorporate the non-linear effects. For phase measurement (i.e. homodyne detection of the $\phi$-quadrature \cite{wiseman:1993a}), we find that the state of the TLS conditioned on the record $\bar J(t)$ obeys the stochastic master equation (SME)
\begin{equation}
	\label{eqn:conditional_qubit_master_equation}
	\begin{split}
	\raisetag{15pt}
		\dot\qrho_{\bar J}\trans{\atD} &= \sL\trans{\atD} \qrho_{\bar J}\trans{\atD} + \sqrt{\Gamma_\mathrm{ci}(t)} \sM[\sz]\qrho_{\bar J}\trans{\atD}(t) 
		[ \bar J(t) - \sqrt{\Gamma_\mathrm{ci}(t)}\mean{\sz}_t ]
		 \\
		&\quad -i\frac{\sqrt{\Gamma_\mathrm{ba}(t)}}{2} \com{\sz}{\qrho_{\bar J}\trans{\atD}(t)}
		[\bar J(t) - \sqrt{\Gamma_\mathrm{ci}(t)}\mean{\sz}_t ]
	\end{split}
\end{equation} 
where $\sL\trans{\atD} \cdot$ is given by \eq{eqn:ReducedMasterEquation} and
		${\cal M} [c]\crho= (c -\mean{c}_t) \crho/2+\crho(c-\mean{c}_t)/2$
is the measurement superoperator with $\mean{c}_t = \mathrm{Tr}[c\crho_J\trans{\atD}(t)]$. The measurement result $\bar J(t)$ is 
	$\bar J(t) = \sqrt{\Gamma_\mathrm{ci}} \mean{\sz}_t + \xi(t)$,
where $\xi(t)$ is Gaussian white noise satisfying $E[\xi(t)]=0$ and $E[\xi(t)\xi(t')]=\delta(t-t')$.   

The SME \eqref{eqn:conditional_qubit_master_equation} is of \ito form and represents a gradual projective measurement of $\sigma_z$.  The rate at which information about the state of the TLS comes out of the cavity is given by
	$\Gamma_\mathrm{ci}(t) =\eta\Gamma_\mathrm{m}\cos^2(\theta_\mathrm{m})$,
where 
\begin{subequations}
	\label{eqn:Gamma_theta_m}
	\begin{align}
		\label{eqn:Gamma_m}
		\Gamma_\mathrm{m} &= \kappa |\beta|^2\left(1 +\frac{|\mu|\cos(\theta_\beta-\theta_\mu)}{4|\beta|n_\mathrm{crit}}+ \frac{|\mu|^2}{64|\beta|^2n^2_\mathrm{crit}}\right),\\
		\label{eqn:theta_m}
			\theta_\mathrm{m}& =\phi-\theta_\beta+
			 \ImaginaryPart\left\{\log\left[1+\frac{|\mu|e^{i(\theta_\beta-\theta_\mu)}}{8 |\beta|n_\mathrm{crit}}\right]\right\},
	\end{align}
\end{subequations}
with $\beta = \ae-\ag$, $\mu=\ae+\ag$, the angles $\theta_\beta = \arg(\beta)$,  $\theta_\mu = \arg(\mu)$ and where $\phi-\theta_\beta \in [0,\pi/2]$.  Furthermore, $\eta$ is a detection efficiency parameter. Choosing the reference phase $\phi$ (the phase of the local oscillator) such that $\theta_\mathrm{m}=0$, the rate of information gain about the TLS state is $\eta\Gamma_\mathrm{m}$. To first order in $\lambda$, this can simply be understood by noting that $\eta\kappa$ is the rate at which photons leak out of the cavity and are detected, while $|\beta|^2$ is the amount of information about the TLS state encoded in the photons. The second order term in $\Gamma_\mathrm{m}$ is a correction arising from the dressing of the TLS by the photons.
Finally,  the rate
	$\Gamma_\mathrm{ba}(t) = \eta\Gamma_\mathrm{m}\sin^2(\theta_\mathrm{m})$
in the SME is extra non-Heisenberg back-action due to the measurement. This rate is zero when $\theta_\mathrm{m}=0$ and is maximum when measuring in the opposite quadrature.  

Using the quantum trajectory equation, the TLS state localizes on one of the basis states as the measurement amplitude is increased~\cite{gambetta:2008a}.  If this localization is faster than the relaxation time $\gamma_1^{-1}$, then the measurement result is a faithful representation of the initial TLS state.   If the TLS has localized in its excited state we do not expect a single jump to the ground state with mean jump time $1/\gamma_1$.  This is in contrast to the predictions from the linear dispersive model~\cite{blais:2004a,gambetta:2008a}.  Instead, telegraph noise due to the measurement-induced excitation rate $\gu$ is expected.  These predictions can be experimentally tested by measuring the waiting time between jumps and comparing to $\gu$ and $\gd$.


As is clear from the above results, increasing the measurement amplitude opens a new excitation channel $\gu$ which leads to a loss of the quantum non-demolition (QND) character of the dispersive measurement.  This in turn will affect the expected SNR of a homodyne measurement of the field.  Indeed, instead of scaling with power~\cite{blais:2004a,gambetta:2008a}, we expect the SNR to saturate with increasing measurement power.  To demonstrate this in the situation where bifurcation is not important, we define the SNR
as
	$\mathrm{SNR} = (\mathrm{SNR}_e + \mathrm{SNR}_g)/2$,
with
	$\mathrm{SNR}_{e(g)} = \Gamma_\mathrm{ci}/(\gamma_{\uparrow e(g)}+\gamma_{\downarrow e(g)})$
for the TLS initially in the excited (ground) state. The rates $\gamma_{\uparrow e(g)}$ and $\gamma_{\downarrow e(g)}$ depend on the TLS state through photon population which is evaluated from the steady-state value of Eq.~(\ref{eqn:Condition_alphas}).

Fig.~\ref{fig:snr}a) shows a plot of the $\mathrm{SNR}$ (solid blue line) as a function of measurement power, $\epsilon_m^2$, and scaled to $n_\mathrm{linear}/n_\mathrm{crit}$ where $n_\mathrm{linear}=\epsilon_m^2/{(\kappa^2/4 +\chi^2)}$ is the average photon number predicted by the linear model.  For concreteness, the parameters chosen are taken from circuit QED~\cite{schreier:2007a}, but these results apply to all cavity QED realizations reaching the strong coupling regime.  Contrary to the result obtained in the linear model (dotted red line), the SNR saturates with increasing power.  For $\kappa/\gamma_1$ small, transients can further reduce the SNR.  This reduction is small for the chosen parameters (dashed-dotted black line).  The value at saturation is within a factor of two or three larger than observed experimentally~\cite{houck_private} for a transmon qubit~\cite{koch:2007a}.  However, at fixed $g$, the ac-Stark shift per photon $2\chi$ for the transmon is reduced substantially by the presence of extra levels resulting in a lower SNR.  Considering the fact that $\eta$ is highly dependent on parameters that are difficult to extract experimentally (e.g.~number of thermal photons produced by the leading amplifier), the agreement with theory is excellent.

The saturation of the SNR is due to the increase in $\gu+\gd$ and to a reduction of the cavity pull with photon number.  Numerical simulations have shown the reduced dispersive ME Eq.~(\ref{eqn:ReducedMasterEquation}) to be very accurate up to about $n_\mathrm{crit}\sim25$ photons.  However, as shown in Fig.~\ref{fig:snr}b) cavity pull in the non-linear model (full blue line) starts to deviate significantly from the exact result obtained from diagonalization of the Jaynes-Cummings Hamiltonian (dashed purple line) at about $n_\mathrm{crit}/2\sim13$.  Our predictions for the SNR therefore underestimate slightly the exact result in the range shown in Fig.~\ref{fig:snr}a).

As shown in Fig.~\ref{fig:snr}c), the SNR is maximized by choosing a measurement frequency such that $\Delta_{rm}'=0$.  This value was used in Fig.~\ref{fig:snr}a).  For this optimal detuning and assuming that $O(n_{\rm crit}^{-1})$ terms can be neglected in Eqs.~(\ref{eqn:Condition_alphas}) and (\ref{eqn:Gamma_theta_m}), we find
\begin{equation}
		\label{eq:SNRapprox}
		\mathrm{SNR}\approx\frac{4 \eta n\kappa (\chi + 2\zeta n)^2 }{[\kappa^2/4+(\chi + 2\zeta n)^2][\gamma_\uparrow  + \gamma_\downarrow]},
\end{equation} 
where we have taken $\phi = \theta_\beta$.  This is shown as the green dotted line in Fig.~\ref{fig:snr}a) where it is seen to be a very good approximation at low power.  

Using these results, it is possible to find parameters that maximize the SNR.  For example, for a given value of $n/n_\mathrm{crit}$ one can find an optimal value of $\kappa$.  If $\gamma_\uparrow  + \gamma_\downarrow$ was independent of $n$, we find from Eq.~(\ref{eq:SNRapprox}) that the optimal value would be $\kappa_\mathrm{opt}/2 = (\chi + 2\zeta n)$ which depends on power.  In general, it will also depend on the relaxation $\gamma_1$ and dephasing $\gamma_\varphi$ rates.  In the case where $\gamma_\varphi\approx\gamma_1$, the optimal $\kappa$ is approximately $\kappa_\mathrm{opt}$, as $\gamma_\uparrow  + \gamma_\downarrow$ is only weakly dependent on $n$.  By contrast, when dephasing dominates over relaxation the optimal $\kappa$ increases with power.  This is to avoid unwanted TLS mixing by limiting the photon population.  Clearly, dephasing plays an important role in the reduction of the SNR and is therefore a crucial parameter to suppress.  For superconducting qubit realizations, efforts in this direction~\cite{koch:2007a} have already paid off~\cite{schreier:2007a}, something which is promising in light of the current results.

In summary, we introduced the dressed dephasing model of dispersive cavity QED.  We obtained reduced master and quantum trajectory equations  which incorporate non-linear corrections.  Dressing by the field leads to mixing rates for the TLS that are proportional to photon number and pure TLS dephasing.  This dressing of dephasing reduces the QND character of dispersive measurements and leads to saturation of the SNR with measurement power.  This is in contrast to earlier studies and is consistant with experimental observations.  These results apply to all physical realizations of cavity QED reaching the strong coupling regime and offer an approach to optimize the SNR  (or measurement fidelity) in the ubiquitous TLS-harmonic oscillator system.

We thank J.~Koch, A.~A.~Houck, R.~J.~Schoelkopf and S.~M.~Girvin for discussions.
MB was supported by NSERC, 
AB by NSERC, FQRNT, CIFAR 
and JG  by MITACS, ORDCF, 
Australian Research Council (CE0348250), and the State of Queensland.


\begin{thebibliography}{10}
\newcommand{\enquote}[1]{``#1''}

\bibitem{haroche:2006a}
S.~Haroche and J.-M. Raimond, \emph{Exploring the Quantum: Atoms, Cavities, and
  Photons} (Oxford University Press, Oxford, 2006).

\bibitem{gleyzes:2007a}
S.~Gleyzes {\it et al.}, Nature \textbf{446}, 297 (2007).

\bibitem{guerlin:2007a}
C.~Guerlin {\it et al.}, Nature \textbf{448}, 889 (2007).

\bibitem{boca:2004a}
A.~Boca {\it et al.}, Phys, Rev. Lett. \textbf{93}, 233603 (2004).

\bibitem{brennecke:2007a}
F.~Brennecke {\it et al.},  Nature \textbf{450}, 268 (2007).

\bibitem{maunz:2005a}
P.~Maunz {\it et al.}, Phys. Rev. Lett. \textbf{94}, 033002 (2005).

\bibitem{hennessy:2007a}
K.~Hennessy {\it et al.}, Nature \textbf{445}, 896 (2007).

\bibitem{englund:2007a}
D.~Englund {\it et al.},  Nature \textbf{450}, 857 (2007).

\bibitem{sillanpaa:2007a}
M.~A. Sillanpaa {\it et al.}, Nature \textbf{449}, 438
  (2007).

\bibitem{wallraff:2004a}
A.~Wallraff {\it et al.}, Nature \textbf{431}, 162
  (2004).

\bibitem{chiorescu:2004a}
I.~Chiorescu {\it et al.}, Nature \textbf{431}, 159 (2004).

\bibitem{johansson:2006}
J.~Johansson {\it et al.}, Phys. Rev. Lett. \textbf{96}, 127006 (2006).

\bibitem{armour:2002a}
A.~Armour {\it et al.},  Phys. Rev. Lett. \textbf{88}, 148301  (2002).

\bibitem{irish:2003a}
E.~K.~Irish {\it et al.},  Phys. Rev. B \textbf{68},  155311 (2003).

\bibitem{schuster:2007a}
D.~I. Schuster {\it et al.}, Nature \textbf{445}, 515 (2007).

\bibitem{thompson:2008a}
J.~D. Thompson {\it et al.}, Nature \textbf{452}, 72 (2008).

\bibitem{blais:2004a}
A.~Blais, {\it et al.}, Phys.  Rev. A \textbf{69}, 062320 (2004).

\bibitem{gambetta:2008a}
J.~Gambetta {\it et al.}, Phys.  Rev. A  \textbf{77}, 012112 (2008).

\bibitem{houck_private}
A.~A.~Houck, B.~R.~Jonhson and R.~J.~Schoelkopf, private communication.

\bibitem{gambetta:2007a}
J.~Gambetta {\it et al.}, Phys.  Rev. A \textbf{76},  012325 (2007).

\bibitem{boulant:2007a}
N.~Boulant {\it et al.},  Phys.  Rev. B \textbf{76}, 014525 (2007).

\bibitem{lupascu:2007a}
A.~Lupascu {\it et al.}, Nature Physics \textbf{3}, 119 (2007).

\bibitem{carmichael:1993a}
H.~J. Carmichael, \emph{An Open System Approach to Quantum Optics} (Springer,
  Berlin, 1993).

\bibitem{gambetta:2006a}
J.~Gambetta {\it et al.}, Phys. Rev. A \textbf{74}, 042318 (2006).

\bibitem{wiseman:1993a}
H.~M. Wiseman {\it et al.}, Phys.  Rev. A \textbf{47}, 642 (1993).

\bibitem{schreier:2007a}
J.~A. Schreier {\it et al.}, arXiv.org:0712.3581v1 (2007).

\bibitem{koch:2007a}
J.~Koch {\it et al.}, Phys.  Rev. A \textbf{76}, 042319 (2007).

\end{thebibliography}

\end{document}